\documentclass[prl,twocolumn,aps,amsmath,nofootinbib,superscriptaddress]{revtex4}
\usepackage{graphicx}
\usepackage{bm}
\usepackage{epsfig}

\newif\ifpdf
\ifx\pdfoutput\undefined
\pdffalse 
\else
\pdfoutput=1 
\pdftrue
\fi

\def\Slash#1{{#1\!\!\!\slash}}

\def\Bslash{B\!\!\!\!\slash}
\def\Dslash{D\!\!\!\!\slash}

\def\bnslash{\bar n\!\!\!\slash}

\def\OMIT#1{}

\newcommand{\nn}{\nonumber} 

\newcommand{\bn}{{\bar n}}
\newcommand{\bea}{\begin{eqnarray}}
\newcommand{\eea}{\end{eqnarray}}

\newcommand{\bnP}{\bar {\cal P}}

\newcommand{\cP}{{\cal P}}

\newcommand{\mcdot}{\!\cdot\!}

\begin{document}
\ifpdf
\DeclareGraphicsExtensions{.pdf, .jpg}
\else
\DeclareGraphicsExtensions{.eps, .jpg}
\fi


\title{
Factorization, Power Corrections, and the Pion Form Factor}
\author{Ira Z. Rothstein}
\affiliation{Dept. of Physics, Carnegie Mellon University, Pittsburgh, PA 15213\footnote{Permanent address}}\affiliation{ 
Dept. of Physics, University of California at San Diego, La Jolla  , 
    CA 02093 
}

\begin{abstract}

This paper  is an investigation of the pion form factor utilizing  recently developed effective field theory techniques. The primary results reported  are: Both the transition and electromagnetic form factors are corrected  at order $\Lambda/Q$ due to time ordered products which account for deviations of  the pion from being  a state composed purely of highly energetic collinear quarks in the lab frame. 
The usual higher twist wave function corrections contribute only at order $\Lambda^2/Q^2$, when the quark mass vanishes.
In the case of the electromagnetic form factor the $\Lambda/Q$  power correction is enhanced by a power of $1/\alpha_s(Q)$ relative to the leading order result of Brodsky and Lepage, if the scale $\sqrt{\Lambda Q}$ is non-perturbative. This enhanced correction could explain the discrepancy with the data.\end{abstract}

\maketitle
\section{Introduction}
Making predictions for hadronic observables is extremely difficult given the complexity of the theory of strong interactions (QCD). In the absence of a solution to the theory, we are forced to accept  reduced predictive power. We can make  predictions only after we have extracted some crucial related information from the data. Moreover, the necessary information is usually in the form of non-local matrix elements and not just fixed couplings.
Even with these lowered expectations, it is still highly non-trivial to find observables which we can predict from first principles. The primary tool at our disposal is factorization\cite{review}. Observables which are ``factorizable" can be separated into long and short distance contributions. Asymptotic freedom allows for a calculation of the short distance piece via a perturbative expansion in the coupling, while the long distance piece is in principle calculable, but in practice must be extracted from experiment. The predictive power lies in the fact that factorization implies universality. That is, the same non-perturbative factor can appear in predictions for disparate processes.

Proving that a certain observable is factorizable is a highly non-trivial process.  The pioneering works on the subject \cite{Lepage, Efremov,Duncan} were based upon diagrammatic techniques and can be quite intricate. Using these techniques, factorization was shown to leading in $1/Q$, where $Q$ is the large energy scale in the process.  For example, the photon-pion transition form factor 
which is defined via ($q=p_\pi-p_\gamma$)
\begin{equation}
\langle \pi^0(p_\pi) \mid J_\mu(0)\mid \gamma(p_\gamma,\epsilon) \rangle=-ie F_{\pi\gamma}p_\pi^\nu \epsilon^\rho q^\sigma \epsilon_{\mu \nu \rho \sigma},
\end{equation}
can be written to leading order (in $1/Q=1/\sqrt{\mid{q}^2\mid}$) as
\begin{eqnarray} \label{LOtFF}
  F_{\pi\gamma} 
  =\frac{2f_\pi}{Q^2} \int_0^1\!\! dx\: C_1(x,Q/\mu)
  \: \phi_{\pi}(x,\mu) \,,\end{eqnarray}
while the pion EM  form factor  defined as
\begin{equation}
\langle \pi(p) \mid J_\mu(0)\mid \pi(p^\prime) \rangle=F_\pi(Q^2)(p_\mu+p^\prime_\mu)
\end{equation}
may be written as
\begin{eqnarray}
F_{\pi}\!=\! (Q_u\!-\!Q_d)\, \frac{f_\pi^2}{Q^2}\, 
\int_0^1\!\!\mbox{d}x\!
~\mbox{d}y \: T_1(x,y,Q/\mu) \phi_\pi(x,\mu) \phi_\pi(y,\mu)\,.\nonumber \\
\end{eqnarray}
In these expressions $\phi_\pi$ carries the universal,  non-perturbative, information about the structure of the pion and is defined via
\begin{eqnarray} \label{pimom}
&&\!\!\!\!\!\!\!\!\!\!\!\!\!\!\!\!-{i}f_\pi\,\delta^{ab}
   \phi_\pi(x,\mu)=\\ \nn & & \!\!\!\!\!\!\! \int \frac{dz}{2 \pi}\, e^{-i x z {\bn\mcdot p} } ~\big\langle ~\pi_{n,p}^a~ \big| 
  ~\bar{q}(z)~ \Gamma_\pi^b W(z,-z)~ q(-z) \big|~ 0~
  \big\rangle,
\end{eqnarray}
where $W$ is a light-like Wilson line needed for manifest gauge invariance.
 $C_1$ and $T_1$ are the perturbatively calculable, process dependent, high energy Wilson coefficients. At asymptotically large values of $Q$, the pion wave function is dominated by its first moment and 
approaches the form $\phi_\pi(x)\propto x(1-x)f_\pi$. Experimentally, the prediction for the transition form factor agrees at the ten percent level, while the EM form factor is off by a factor of order one. Thus, the relevant questions is,  
how large are the subleading corrections in each case? 
This paper undertakes the task of systematically categorizing  the power corrections to these processes, including the contribution from 
the so-called ``end-point region" . The issue of the breakdown of factorization
will also be discussed.

\section{SCET}
The results in this paper are derived  utilizing recent developments in the so-called Soft-Collinear Effective Theory (SCET)\cite{SCET}. In this approach, factorization proofs simplify because modes with varying virtualities are separated at the level of the Lagrangian\cite{bfprs}. Proving factorization becomes tantamount to determining if the theory properly accounts for the IR physics  of  the process under consideration. 
Perhaps the true power of the effective field theory approach, as applied to  exclusive processes,  is that it enables one to account for  power corrections in a systematic fashion including the so-called ``end-point" contributions\footnote{These contributions are sometimes also referred to as the "Feynman" or "soft" regions.}.  

To build the proper effective field theory  one isolates the degrees of freedom responsible for the non-analytic behavior of the low energy theory. Usually this corresponds to simply integrating out massive fields, but there are cases \cite{LMR} where one wishes to explicitly separate  certain subset of fluctuations of the fields. This separation allows for  manifest power counting, which
 in turn makes the classification of power corrections relatively simple. In SCET the relevant modes in the IR are:  
collinear $p_c = (p_c^+, p_c^-, p_c^\perp)=(n\mcdot p_c, \bn \mcdot p_c,
p_c^\perp) \sim Q(\lambda^2, 1,
\lambda)$, soft $p_s \sim Q(\lambda, \lambda, \lambda)$ and usoft $p_{us} \sim
Q(\lambda^2, \lambda^2, \lambda^2)$, where $n^2=\bn^2=0$, $n\mcdot\bn=2$, and
$\lambda \ll 1$ is the expansion parameter. The relevant modes are fixed by the external momenta and use of the Coleman-Norton theorem. In \cite{bfprs}, the authors chose  $\lambda\propto \Lambda/Q$, as this fixes the transverse momenta of the external lines to be of order $\Lambda$, as it should be physically. This also means that the usoft modes have virtuality less than $\Lambda$, that is,
their wavelength is longer than the confinement radius. These "hyper-confining modes" should not
contribute to physical processes. In all the cases discussed in \cite{bfprs}, it was shown that
these modes do indeed cancel in the matrix elements. There is one additional mode which
may contribute in the case when there are at least two collinear directions. Namely, the so-called
``Glauber region", whose momenta scale as $(\lambda^2,\lambda^2,\lambda)$. While these modes
typically are not relevant in exclusive processes, this has  yet to been shown within the context of
SCET.

 \subsection{The Need for an Intermediate Theory}
 
 New subtleties in the theory arise when one considers power corrections to exclusive processes due 
 to what is known as the ``end-point" region, which is defined as the contribution to the process
 where the hadron is in an asymmetric configuration. That is, one or more of the constituents
 carries a parametrically small part of the longitudinal momentum \footnote{I will always be
 discussing processes/frames where the hadron is nearly light-like.}.
 In the context of effective field theory language, this configuration corresponds to a sub-leading
 fluctuation. To see this note that to preserve manifest power counting, we take the pion to be 
 an eigenstates of the leading order Hamiltonian of the effective theory, which as will be discussed
 below, only contains collinear fields. As such, the overlap of the pion with a field  which contains a net soft quantum number \footnote{That is, an operator
 which has no overlap with a purely collinear state.} will be nil. Thus, in the effective field
 theory  formalism these soft regions of the wave function show up via time ordered products
wherein  a collinear field fluctuates to a soft field and then back to a collinear field. 
 These time ordered products must involve an operator which couples a collinear
 field with a field whose momenta scale as $(\Lambda,\Lambda,\Lambda)$. However, 
 momentum conservation forbids the coupling of such a field with a collinear field
 whose momentum scales as ($Q,\Lambda^2/Q,\Lambda$). 
 
 This issue was addressed in \cite{bps3}, where it was pointed out that such interactions
 will be properly accounted for, if one considers working in two stages. In the initial theory,
 labeled SCETI, the scaling parameter $\lambda$ is order $\sqrt{\Lambda/Q}$, this theory
 is valid at scales below $Q$ but above $\sqrt{\Lambda Q}$.  In this theory interactions
 between collinear $(1,\lambda^2,\lambda)$ and ultra-soft ($\lambda^2,\lambda^2,\lambda^2$)
 gluons are permitted. At the scale $\sqrt{\Lambda Q}$, we match onto a second
 theory SCETII, where $\lambda$ now scales as $\Lambda/Q$. In doing this matching
 we integrate out modes with invariant mass $q^2\geq \Lambda Q$. In SCETII, there 
 are no interactions between collinear and soft modes\footnote{One can think of the ultra-soft
 modes in SCETI as becoming the soft modes in SCETII.}
  
 \section{The Transition Form Factor}

\subsection{Leading Order}
In the effective theory
each mode is interpolated by a distinct field  and  scales homogeneously in $\lambda$. For instance, collinear modes with large light-cone momentum in the $n$ direction are interpolated by $\xi_n$ and $A_n^\mu$ for fermions and gauge bosons respectively.
These fields have support only over momenta of order $\lambda^2$, as their large light cone and transverse momenta have been scaled out.
 The leading order Lagrangian for these collinear fields with momenta 
in the $n$ direction, is given by
\begin{eqnarray} \label{L0}
 && {\cal L}^{(0)}_c = \bar \xi_n \left[ i n \mcdot D 
  + i \Slash{D}^c_\perp \frac{1}{i \bn \mcdot D_c}  i \Slash{D}^c_\perp\right]  
  \frac{\bnslash}{2} \xi_n + {\cal L}^{(0)}_{cg} \,,\hspace{0.9cm}
\end{eqnarray}with 
\begin{eqnarray}
i\bn\mcdot D_c&=&\bnP\!+\!g\bn\mcdot A_n, \nonumber \\
 iD_c^\perp&=&\cP^\perp\! +
\!gA^\perp_n \nonumber \\
in\mcdot D&=&in\mcdot\partial+gn\mcdot A_{us}+gn\mcdot A_n.
\end{eqnarray}
The operators $\bnP$ and $\cP^\perp$ are derivative like operators whose eigenvalues are the large light cone and transverse momenta respectively. 
 The 
gluon action, ${\cal L}^{(0)}_{cg}$, can be found in \cite{SCET}.

Let us start by considering the pion transition form factor. 
The leading order matching for this process was performed in \cite{bfprs}.
This process involves the scattering of a highly virtual
photon and a quark-anti-quark constituent pair  of an on-shell photon in the $n$ direction. 
We will work to leading order  in $\alpha_{em}$.
\footnote{If we were to treat the photon like a hadron, then the analysis is almost identical to the case of the EM form factor discussed in later paragraphs.}
  By integrating out the hard off-shell intermediate state in the $\gamma^\star +\gamma\rightarrow q+\bar{q}$ process, we generate a two quark operator in the SCETI.
The
most general spin structures for currents with two collinear particles
moving in the same or opposite directions are \cite{bfprs}

\begin{eqnarray} \label{Jspin}
   \bar\xi_{n}\,\big\{ \bnslash\,, \bnslash\gamma_5\,,
                 \bnslash\gamma^\mu_\perp \big\} \xi_{n}~~~;~~~
     \bar\xi_{\bn} \big\{ 1 \,, \gamma_5 \,, \gamma^\mu_\perp \big\}  \, \xi_{n}.\end{eqnarray}
From this result we can see that, for the case at hand, there is only one relevant operator structure which interpolates for the pion namely $\bnslash \gamma_5$, thus the leading order matching result is of the form
\begin{eqnarray} \label{Opig}
  O^{(0)}_{\omega_1,\omega_2}&=& i\epsilon^\perp_{\mu \nu} (\bar{\xi}_{n}W)_{\omega_1}\, \bar n \!\!\! \slash \gamma_5
    (W^\dagger \xi_{n})_{\omega_2}~, 
\end{eqnarray}
where the isospin structure has been suppressed, and $\epsilon^\perp_{\mu \nu}=\frac{1}{2} \bar{n}^\rho n^\sigma\epsilon_{\mu \nu\rho \sigma}$.
Here the $W$'s are the Fourier transforms of light-like Wilson lines 
\begin{equation}
W(y,\infty)=P exp\left(i\int_y^\infty \bar{n} \cdot A_n(\lambda \bar{n}) d\lambda\right),
\end{equation}
and $\omega_{1,2}$ the total collinear momenta of the jet like structure $W^\dagger \xi$ and $\bar{\xi} W$ respectively.
Typically each operator is accompanied by label subscripts, such labels are implied if they are not
explicit.
There is an implied sum over all label momenta such as $\omega_{1,2}$, which will be restricted
by momentum conservation when we take a matrix element.
We may then decouple the usoft modes from the collinear modes in the Lagrangian via field redefinitions [5]. \begin{eqnarray}\label{redef}
 \hat{\xi}_{(n,\bar{n})} = Y_{(n,\bar{n})}^\dagger \xi_{(n,\bar{n})}\,, \quad 
 \hat{A}_{(n,\bar{n})} = Y_{(n,\bar{n})}^\dagger A_{(n,\bar{n})} Y_{(n,\bar{n})}~, 
\end{eqnarray}
where $Y$ is an usoft Wilson line
defined as
\begin{equation}
Y_{n}=P exp\left(i\int_y^\infty n \cdot A(\lambda n) d\lambda\right).
\end{equation}

 This redefinition has the effect of decoupling usoft lines from collinear lines in the action, at the cost of introducing $Y$ factors into the operator $O_0$. However, we can see that since the $Y$ are usoft and carry no large light-cone momenta they will cancel in $O_0$ as a consequence of unitarity.  The $Y$'s will however, show up  in sub-leading Lagrangian and external operators.
 I will drop the
hatted symbol from here on, and thus the reader should assume that all fields have been redefined as in (\ref{redef}).

The matching onto SCETII is trivial at this order. The off-shellness of the external collinear lines is reduced to being less than $\Lambda Q$.
Taking the matrix element of this operator between the vacuum and one pion state yields the usual leading order result in terms of the pion wave function~(\ref{LOtFF}).

\subsection{Power Corrections}
 Power corrections 
 arise from either matching onto higher order operators or from including corrections from the sub-leading Lagrangian into time ordered products (TOP's).
The order $\lambda$ action introduces couplings between usoft and collinear fields and is given by
~\cite{chay,beneke,rpi}
\begin{eqnarray}
&& {\cal L}_{\xi\xi}^{(1)} = \bar \xi_n  i \Slash{D}^{us}_\perp 
  \frac{1}{i \bn \mcdot D_c}  i \Slash{D}^c_\perp \frac{\bnslash}{2} \xi_n
  \mbox{ + h.c.}\,,  \\
&& {\cal L}_{cg}^{(1)} = \frac{2}{g^2} {\rm tr} 
  \Big\{ \big[i {\cal D}^\mu , iD_c^{\perp\nu} \big] 
     \big[i {\cal D}_\mu , iD_{us\,\nu}^\perp \big] \Big\} + {\rm g.f.}\,, \nn \\
    && {\cal L}^{(1)}_{\xi q} = ig\: \bar\xi_n \: \frac{1}{i\bn\mcdot D_c}\: 
 \Bslash_\perp^c W  q_{us} \mbox{ + h.c.}\,. \nn
\end{eqnarray}
with ${\cal D}^\mu = n^\mu \bn\mcdot D_c/2 + D_c^{\perp\mu} +\bn^\mu n\mcdot
D/2$, g.f.~denotes gauge fixing terms, and 
\begin{equation}
igB_{\mu \perp}^ c \equiv [i\bar{n} \cdot D_c,iD_{\mu c}^\perp]\equiv\bn_\nu ({
G}_{n})^{\nu\mu_\perp}.
\end{equation}
 The collinear gauge invariant field strength is
\begin{eqnarray} \label{covG}
  ({G}_{n})^{\mu\nu} = -\frac{i}{g}\Big[
   [i{\cal D}_n^\mu + gA_{n,q}^\mu, 
i{\cal D}_n^\nu+gA_{n,q'}^\nu ] 
   \Big] \,.
\end{eqnarray}

We will also need the order $\lambda^2$ Lagrangian
\begin{eqnarray} \label{L12}
{\cal L}^{(2)}_{\xi \xi}\!\!\! &=& 
\!\!\!\bar \xi_n \!\!\left(\!
\Dslash_{us}^\perp \frac{i}{\bn\cdot D_c} \Dslash_{us}^\perp
 \!-\!
\Dslash^\perp_c \frac{i}{\bn\cdot D_c}
\bn\cdot D_u \frac{1}{\bn\cdot D_c} \Dslash^\perp_c
\! \!\right)\!
\frac{\bnslash}{2}\xi_n.\nn \\
\end{eqnarray} 
 As discussed above, it is important to understand that  time ordered products corrections correspond to perturbations of  the states in the effective theory. That is, in the effective theory the pion state is not the physical pion state, it contains only collinear modes in SCETII. The true pion is an eigenstate of the full Hamiltonian, so including perturbations into the time ordered product accounts for this difference in a systematic fashion.  There is a direct analogy with HQET \cite{MW} which is perhaps illuminating.
 In the limit where the quark mass is taken to infinity, the B meson is completely static.
 It is true that in the physical meson the quark has some kinetic energy, but these effects
 can be included when considering power corrections, via time ordered products with
 sub-leading operators.  In this way we can build up the full meson state order by order in the
 inverse quark mass.
  For a pedagogical discussion on the subject see \cite{TASI}.

 Matching onto SCETI we may generate order $\lambda$ operators by inserting  $B ^{c \perp}_\mu$ into $O^{(0)}$,
\begin{equation}
O^{(1)}=i\bar{n}_\nu \epsilon_{\perp_{\mu \rho}}(\bar \xi _n  W)_{\omega_1}\bar{n}\!\!\! \slash  \gamma_5
 (W^\dagger B ^{\rho \perp}_nW)_{\omega_2} (W^\dagger \xi_n)_{\omega_3}~,
\end{equation}
 But this operator does not interpolate for the pion \footnote{We work in a frame where the transverse momentum of the pion vanishes.}. Operators with insertions of $D^c_\perp$, can be absorbed into $O^{(1)}$.
 We may also consider quark mass effects, 
which may arise from either the expansion of the fields via
\begin{equation}
q=\left(1+\frac{1}{\bn \cdot iD_c}(iD^c_\perp \!\!\!\!\!\!\!\slash ~-m_q)\frac{\bnslash}{2}\right) \xi_n, 
\end{equation} 
 or from including  mass effects in perturbative matching.
Using a simple spurion\cite{text} analysis and the fact that there is only one possible non-vanishing Dirac structure in the effective theory which violates chirality, it is simple to show that matching cannot generate any $O(\lambda^0)$ operators \footnote{Though the operator itself is leading order it would come in suppressed by $m_q/Q$.} which are linear in the quark mass to all orders in perturbation theory. 
However, the introduction of a quark mass  generates a new $O(\lambda)$ operator
\begin{equation}
O^{(1)}_m\!\!=\!\!\ i\epsilon^\perp_{\mu \nu} (\bar \xi_n W)_{\omega_1}\epsilon^\perp_{\alpha \beta}(W [ i\overleftarrow{D}^\alpha_\perp \gamma^\beta_\perp -i\overrightarrow{D}^\alpha_\perp 
\gamma^\beta_\perp] W^\dagger)_{\omega_2}
(W^\dagger\xi_n )_{\omega_3}.
\end{equation}
The spontaneous breaking of chiral symmetry would lead these contributions to be 
numerically enhanced.

The possible order $\lambda^2$ operators which are bi-linear in the collinear quarks are \begin{equation} \label{loc}
O^{(2)}_a= i\epsilon^\perp_{\mu \nu}(\bar \xi_n W)_{\omega_1} (W^\dagger_n n \mcdot D W^\dagger )_{\omega_2}  \bnslash\gamma_5(W^\dagger \xi_n)_{\omega_3}  
\end{equation}
\begin{equation}
O^{(2)}_b\!\!\!= \!\!  i\epsilon^\perp_{\mu \nu}(\bar \xi_n W)_{\omega_1} (W \Slash{D}^c_\perp W^\dagger)_{\omega_2}  (W\Slash{D}^c_\perp W^\dagger)_{\omega_3}
 \bnslash\gamma_5(W^\dagger\xi_n )_{\omega_4} 
\end{equation}
where the appearance of the momentum subscript implies the existence of non-trivial Wilson coefficients which account for possible insertions of the operator 
$\bn\cdot D_c$.  $ O^{(2)}_b$, is a representative of a class of operators with two transverse covariant derivatives acting in all possible ways. The Wilson coefficients of these operators will be 
related by reparameterization invariance\cite{rpi}. 
Other operators involving collinear field strength operators can be expressed in terms of linear combinations of these operators. 
In addition, it is possible to generate operators with one collinear and one usoft quark,
i.e. $\bar \xi_n W \bar n \!\!\!\slash ~\gamma_5 q_{us}$, but the contribution from this operator will be order $\lambda^3$, since an additional insertion of a sub-leading operator is necessary to get a non-vanishing matrix element between pion and vacuum. 

 We now match onto the lower theory SCETII, where external virtualities are restricted to be less than $\Lambda Q$. In doing so, the external states will pick out a subset of the  collinear modes, whose transverse label momentum is of order $\Lambda$. In addition, the usoft modes get relabeled to be soft modes. For Wilson lines this transformation is denoted by $Y\rightarrow S$.
 $O_a^{(2)}$ will now scale as $O(\Lambda^2/Q^2)$, since $n\cdot D$ picks out the smallest
 component of the collinear field which now scales in this way.
The operators in the class of $O_b^{(2)}$ will have vanishing Wilson coefficients when matching onto SCETII since we set the external transverse momenta to zero. Formally this occurs because
\begin{equation}
\cP^\perp \xi^{II}=0.
\end{equation}
 However, there are a class of identical operators with usoft derivatives as well which will give $\Lambda^2/Q^2$ corrections in SCETII. This scaling arises because usoft modes  in SCETI  match onto soft modes in SCETII, which scale as $\Lambda/Q$.
In addition, we will generate new operators by considering the time ordered products in SCETI
\begin{eqnarray}
T_1&=&\int d^4 x d^4y T\left( O^{(0)} {\cal L}_{i}^{(1)}(x) {\cal L}_{i}^{(1)}(y)\right) \nn \\
T_2&=&\int d^4 x   T\left( O^{(0)} {\cal L}_{j}^{(2)}(x) \right), 
\end{eqnarray}
where ${\cal L}_{i}^{(1)}$ and ${\cal L}_{i}^{(2)}$ are any of the first and second order Lagrangian corrections respectively. Since the external states have vanishing transverse momenta as far as the label operators in SCETI are concerned, any operator with an odd number of transverse derivatives vanishes. TOPs with an even number of derivatives need not vanish since there exists non-vanishing
Wick contractions\footnote{Note this is not true for the ``local'' operators $O_b^{(2)}$.}.
The TOPs $T_1$ and $T_2$ however,  will have non-vanishing matching coefficients and will contribute at order $\Lambda/Q$.

To see how this scaling comes about, note that 
while the power corrections to the Lagrangian  ${\cal L}_{i}^{(1,2)}$ will scale down from 
being order $\Lambda/Q$ to $\Lambda^2/Q^2$, this will not happen within the time ordered
product. The reason for this is that once we lower the virtuality of the external lines down
to $q^2 \sim \Lambda^2$, the collinear field scaling goes from $\sqrt{\Lambda Q}$ to $\Lambda$, but  internal collinear lines connecting operators at distinct points will still scale as they did in SCETI. This leads to enhancements in time ordered products relative to
``local'' operators. For instance, if we consider $T_2$, then a Wick contraction of collinear lines
between the two operators leads to a fermion line which is off-shell by $\Lambda Q$. 
Thus the effective scaling of the product of these two fields is $\Lambda/Q$, whereas if
they carried offshellness $\Lambda^2$, the scaling would be just $\Lambda^2/Q^2$, hence
the enhancement. Furthermore, if the scale $\sqrt{\Lambda Q}$ is perturbative
we may integrate out these off-shell modes perturbatively.


Thus the correction to the leading order result may be written formally as
\begin{eqnarray}
\frac{Q^3}{2i}\delta F_{\pi \gamma}\!\!\!\!&=&\!\!\!\!\!\int\! \!\!d^4 x d^4y\!\!\! \int d\omega_j  C(\omega_j)\langle \pi_{n,p}\!\! \mid \!\! O^{(0)}_{\omega_j} {\cal L}_{i}^{(1)}(x) {\cal L}_{i}^{(1)}(y)\!\!\mid \!\! 0 \rangle\nonumber \\
&+& \int d^4 x \!\! \int d\omega_j  C(\omega_j)\langle \pi_{n,p} \!\mid\!\! O^{(0)}_{\omega_j} {\cal L}_{i}^{(2)}(x)\!\!\mid \!\! 0 \rangle.
\end{eqnarray}
Note that the large light cone momenta ($\omega_i$) flowing out of the leading order operator
will now flow into the vertices of the sub-leading Lagrangian insertions.  

It is interesting to note that in the SCET formalism {\it all} the power corrections are factorable, in the sense that they can be written as products of matrix elements of various types of fields \footnote{I thank Mark Wise and Iain 
Stewart for emphasizing this point to me.}.
That is, factorization is manifest in all the power corrections, since after the field redefinition, none of the various types of fields communicate. As a consequence of this, all of the soft fields may be factored
into 
 vacuum matrix elements which are independent of the hadron, since these fields have no overlap 
 with the hadrons. This implies that, under our working assumption that SCET, in particular SCETII,  as formulated in \cite{SCET,bps3} is the appropriate effective field theory for the above processes, the soft pieces of all hadrons are universal. 
Thus, there is hope that we can extract the soft structure functions and use them to make predictions in disparate processes.

\section{The Electromagnetic Form Factor}
\subsection{Leading Order}
Let us now consider the EM form factor. As we will see, the existence of two jets will have important ramifications. 
In this case,  the Lagrangian splits into two pieces, one for each type of collinear mode, which do not communicate,
\begin{equation}
{\cal L}_{tot}={\cal L}_{\bn}+{\cal L}_{n}.
\end{equation}

The usual leading order Brodsky-Lepage (BL) result
 was regained in SCET by matching the full QCD current onto four quark operators which are generated at order $\alpha_s(Q)$ \cite{bfprs}, 
\begin{eqnarray}O^{(0)}_{1}&=&(n+\bar{n})_\mu \times \nn \\
&& (\bar \xi _n  W_n)_{\omega_1} \bar{n}\!\!\! \slash  \gamma_5 (W^\dagger_n \xi_n)_{\omega_2}(\bar \xi _\bn  W_{\bn})_{\omega_3} n\!\!\! \slash  \gamma_5 (W^\dagger_\bn \xi_\bn)_{\omega_4} \\ \nonumber
O^{(0)}_{8}&=&(n+\bar{n})_\mu \times \nn \\
&&(\bar \xi _n  W_n)_{\omega_1} \bar{n}\!\!\! \slash  \gamma_5 
T^a(W^\dagger_n \xi_n)_{\omega_2}(\bar \xi _\bn  W_\bn)_{\omega_3} n\!\!\! \slash  \gamma_5 T^a(W^\dagger_\bn \xi_\bn)_{\omega_4}.\nn
\end{eqnarray}
The subscripts of these operators denote their color representation.

Scaling the fields by the usoft Wilson lines has no effect.  The fields of opposing directions do not couple and therefore, the matrix element of these operators between pion states of opposing light like directions factorizes into the result (\ref{pimom}), with the octet contribution vanishing. 
\begin{figure}[t]
 \centerline{
  \mbox{\epsfysize=2.2truecm \hbox{\epsfbox{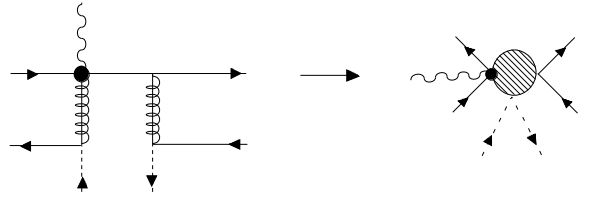}} }
  }
\vskip-0.6cm
\caption[1]{Integrating out modes with virtuality $\sqrt{\Lambda Q}$ lead to operators which represent the ``soft" piece of the wave function. The solid (hatched) circle represents modes off-shell by $Q$ ($\sqrt{\Lambda Q}$).}

\label{fig_FACT} 
\vskip-0.5cm
\end{figure}
\subsection{Power Corrections}
When we're concerned with power corrections, we must consider the matching onto operators
which may not necessarily have non-vanishing overlap with the pions.  A non-zero  overlap
can be achieved via some TOPs' with subleading terms in the Lagrangian.
In the EM  case we may also match onto a quark bilinear 
\begin{equation}
   O^{(-2)}(\omega_1,\omega_2)=(\bar\xi_{\bn}W_{\bn})_{\omega_1} \Gamma 
(W^\dagger_n \xi_n)_{\omega_2} \,,
\end{equation}
where the possible Dirac structures, $\Gamma$ are given in (\ref{Jspin}). 
This operator is enhanced by a factor $\lambda^{-2}$ relative to the four quark operator since each collinear fields scales as $\lambda$.
However, charge conjugation implies that the only structure with non-vanishing Wilson coefficient is $\gamma_\mu^\perp$, which will not contribute for the case of the pion but will for the case of the $\rho$. At next order in the matching the operators 
\begin{equation}
O^{(-1)}_a\!=\! 
\bn_\mu ( \bar\xi_{\bn} 
 \overleftarrow{D}_{c,\bn}\!\!\!\!\!\!\!\!\!\!\!\slash^\perp~~~ W_{\bn})_{\omega_1}(W_n^\dagger \xi_n)_{\omega_2}
 \end{equation}
 \begin{equation}
 O^{(-1)}_b= n_\mu(\bar\xi_{\bn} W_{\bn})_{\omega_3} (W_n^\dagger \overrightarrow{D}_{c,n}\!\!\!\!\!\!\!\!\!\!\!\slash^\perp~~\xi_n)_{\omega_4}
%
\end{equation}
are generated at order $\alpha_s^0$.  Current conservation and RPI will relate the Wilson coefficients
$(C_a,C_b)$ of these two operators such that  $\omega_1 C_a+\omega_4 C_b=0$.
Note that we have introduced here two covariant derivatives with labels $n$ and $\bar{n}$, referring
 to the two distinct collinear sectors.

As opposed to the previous sub-leading operators, $O_{a,b}$ are not invariant under the ultra-soft  field redefinitions
(\ref{redef}). In the redefined basis we have
\begin{equation}
O^{(-1)}_a\!=\! 
\bn_\mu ( \bar\xi_{\bn} 
 \overleftarrow{D}_{c,\bn}\!\!\!\!\!\!\!\!\!\!\!\slash^\perp~~~ W_{\bn}Y_\bn)_{\omega_1}(Y^\dagger_n W_n^\dagger \xi_n)_{\omega_2}
\end{equation}
 \begin{equation}
 O^{(-1)}_b= n_\mu(\bar\xi_{\bn} W_{\bn}Y_\bn)_{\omega_3} (Y^\dagger_{n}W_n^\dagger \overrightarrow{D}_{c,n}\!\!\!\!\!\!\!\!\!\!\!\slash^\perp~~\xi_n)_{\omega_4}.
\end{equation}
The leading order matrix elements between back to back pions of this operator vanish. To generate a non-zero overlap, all that is needed is the proper insertion of sub-leading operators which
will inject one collinear quark into each jet.
Furthermore, to get a non-vanishing Wilson coefficient the TOP should contain an even number of 
insertions of $D\!\!\!\slash_c^\perp$\cite{bps3}.
This can be accomplished via a usoft (which become soft in SCETII) partonic fluctuation. The TOP
\begin{eqnarray}
T_{us}(\omega_i)
&=&\int  \!\!\! d^4 x_i T[(C_a(\omega_1,\omega_2) O^{(-1)}_{a}{\cal L}_{\xi \xi}^{n (1)}(x_1) \nonumber \\
&+&C_b(\omega_3,\omega_4) O^{(-1)}_b{\cal L}_{\xi \xi}^{\bar{n} (1)}(x_1) ){\cal L}_{\xi q}^{\bn (1)}(x_2){\cal L}_{\xi q}^{n (1)}(x_3)] \nonumber \\
\end{eqnarray}
gives an order $\Lambda/Q$ contributions to the EM pion form factor of the form
\begin{equation}
Q^2 \delta  F_{\pi}=\int d\omega_i \langle \pi_{n,p} \mid T_{us}(\omega_i)+h.c. \mid \pi_{\bar{n},p^\prime} \rangle.
\end{equation}
This type of corrections was anticipated in \cite{Lepage}
and is simply a correction to the pion state. 
Note that each of the operators must be accompanied by a subleading Lagrangian 
which contains a collinear gluon moving in the opposite direction relative the the gluon
in the operators. This insertion is needed to ensure that there are an even number of
transverse covariant derivatives in both directions.

If the scale $\Lambda Q$ were perturbative then
we could integrate the intermediate gluons  as depicted in figure 1. The ultra-soft quark lines, being
fluctuations of virtuality $\Lambda^2$, would only be closed once the matrix element is taken. 
There is an additional operator which gives an identical contribution but 
with $n\leftrightarrow \bn$. 
TOPs at this order involving insertions of the quark mass have zero matching coefficients as they involve powers of the external transverse momentum. However, 
there will be quark  mass dependence in the non-perturbative matrix elements in SCETII which will be relevant for $SU(3)$\cite{Wise} and are only down by $m/\Lambda$.


Notice that the usual leading order BL result is proportional to $\alpha_s(Q)$ while this soft contribution scales as $\alpha_s(\sqrt{Q\Lambda})^2$. The data only reaches $Q^2=10~GeV^2$, with error bars as large as the  
signal itself at larger $Q$\cite{emdata}. Given that $\Lambda$ is of order $1$ GeV, the scale $\sqrt{\Lambda Q}$ is likely  non-perturbative over most of the range of the 
data, giving the TOP an enhancement of $1/\alpha_s(Q)$ relative to the leading order BL result.
Thus, we see that the lack of concordance between theory and data may be due to the enhanced power correction in the EM form factor. Of course, there are other possible reasons for the discrepancy. It could be that using the asymptotic wave function is a poor approximation for these values of $Q$.
However, the fact that the transition form factor seems to agree with the 
data\cite{data}, within theory errors, lends credence to the possibility that the enhanced power correction discussed here could be the real culprit. 
Finally, the discrepancy could also be due to the extrapolation of the $\gamma^\star p\rightarrow \pi n$ data to the pion pole\cite{Carlson}.

This work was supported in part by the DOE under grants DOE-ER-03-40682-143 and
DE-AC02-76CH03000.

\end{document}